\documentclass[preprint]{revtex4-1}

\usepackage{amsmath,amssymb}
\usepackage{graphicx}%
\usepackage{bm}%

\begin{document}

\title{Scaling of Temperature Dependence of Charge Mobility\\
in Molecular Holstein Chains}

\author{D.A.~Tikhonov, N.S.~Fialko, E.V.~Sobolev, and V.D.~Lakhno}

%
%
%

\altaffiliation{%
Institute of Mathematical Problems of Biology RAS, Pushchino, Russia.
E-mail: fialka@impb.psn.ru}

\date{\today}

\begin{abstract}
The temperature dependence of a charge mobility in a model DNA based on
Holstein Hamiltonian is calculated for 4 types of homogeneous sequences
It has turned out that upon rescaling all 4 types are
quite similar.  Two types of rescaling, i.e.\ those for low and
intermediate temperatures, are found. The curves obtained are
approximated on a logarithmic scale by cubic polynomials. We believe
that for model homogeneous biopolymers with parameters close to the
designed ones, one can assess the value of the charge mobility without
carrying out resource-intensive direct simulation, just by using a
suitable approximating function.
\end{abstract}



\maketitle

\section{Introduction}

Presently, considerable attention of researchers is focused on
biological macromolecules, such as DNA, which are a promising object
to be used in na\-no\-bio\-elec\-tro\-nics \cite{bib1,bib2}, for example in constructing
electronic biochips and using DNA as molecular wires. The value of
conductivity in the chain can be assessed with the knowledge of charge
mobility and concentration of free charges.

Our calculations are based on Holstein model. Despite its simplicity
this model is widely used for description of DNA charge transport \cite{refer2-1,refer2-2,refer2-3,refer2-4}.
Using the semiclassical Holstein model we calculated a diffusion coefficient $D$ from
which the value of the charge mobility $\mu$ in homogeneous polyG, polyC,
polyA and polyT DNA fragments was found (on the assumption that the
charge formed on one DNA strand cannot jump to the other) for a wide
range of thermostat temperatures $T$. For different chains occurring at
the same temperature $T$, the calculated values of $D$ are obviously
different (the difference between polyA and polyT is nearly two orders
of magnitude). However the temperature dependence of the diffusion
$D(T)$ seems to be alike in all the cases.

It turned out that upon rescaling of $D$ and $T$ by the values
depending on the overlapping integral between neighbouring sites of
the chain, all the graphs $D(T)$ lie very close to one another, on the
interval 100--1000\,K the difference not exceeding 5\%. We believe
that for a homogeneous biopolymer with parameters close to DNA-modeled
ones, one can assess the value of the charge mobility at temperature
$T$ without carrying out model calculations, just by associating it
with a point with appropriate coordinates on the ``rescaled'' graph.

The paper is arranged as follows.
In Section \ref{SecMod} we introduce a semiclassical Holstein Hamiltonian and relevant motion equations
which are modified by Langevin approach in such a way as to involve the terms responsible for the contribution of temperature fluctuations.
In this formulation, the problem is of interest not only for DNA but also for a wide range of one-dimensional molecular systems
in which phonon dispersion is negligible.
In Section \ref{SecParam} we describe an approach for calculating the diffusion coefficient of a quantum particle in a classical molecular chain.
There we present parameter values for homogeneous nucleotide chains used in further calculations.
In Section \ref{SecScale} we give the calculation data obtained for the diffusion coefficient of a hole in homogeneous chains.
It is shown that for these values, one can get universal approximations of their temperature dependencies
in a wide temperature range.
The results obtained can be used for any molecular chains with optical phonons.
In Section \ref{SecAdd} we consider Holstein Hamiltonian with dispersion.
This Hamiltonian immediately stems from the Peyrard--Bishop model in the absence of anharmonicity \cite{bib162}.
Consideration of the chain dispersion in the case of DNA means taking account of the contribution of stacking interaction into its dynamics.
In this case the temperature dependence of the diffusion coefficient also falls on the universal approximating curve obtained for medium temperatures.
In this section we also investigate charge transfer in regular and homogeneous chains with regard to solvation effects.
It is shown that for these chains, the approximation found does not suit.
In Section \ref{SecEnd} we discuss the results obtained.

\section{Model}
\label{SecMod}

Modeling is reduced to solving a system of ordinary differential
equations which describe motion of a fast quantum particle (electron
or a hole) over a chain of classical sites. In order to take account
of the thermostat temperature, classical equations involve terms with
viscous friction and random force possessing special statistical
properties (Langevin equations). Calculations are carried out for a
large number of simulations (i.e., dynamics of charge distribution from
various initial conditions and with various values of random force) so
that to calculate subsequently the values of macroscopic physical
quantities ``averaged over simulations''.

The model is based on Holstein Hamiltonian for a discrete chain of
sites \cite{bib3} (Holstein considered a chain of two-atom sites, in the case
of DNA a complementary nucleotide pair is thought to be a
site \cite{bib4,bib5,bib6}).
In a semiclassical approximation, choosing a wave function $\Psi$ in the
form $\Psi = \sum_{n=1}^N b_n |n\rangle$,
where $b_n$ is the amplitude of the probability of the charge
(electron or hole) occurrence on the $n$-th site ($n=1,{\ldots} ,N$,
$N$ is the chain length), we write down the averaged Hamiltonian:
\begin{align}
\langle \Psi | \hat{H} |\Psi \rangle &=
\sum_{m,n} \nu_{nm} b_m b_{n}^* +
\frac 12 \sum_n M \dot{\tilde{u}}_{n}^2 +
\nonumber \\
&{}+
\frac 12 \sum_n K \tilde{u}_{n}^2 +
\sum_n \alpha' \tilde{u}_{n}  b_{n} b_{n}^* .
\label{eq1}
\end{align}
Here $\nu_{mn}$ ($m\neq n$) are matrix elements of the electron transition
between $m$-th and $n$-th sites (depending on overlapping integrals),
$\nu_{nn}$ is the electron energy on the $n$-th site.
We use the nearest neighbour approximation, i.e.\ $\nu_{mn} = 0$,
if $m \neq n\pm 1$;
suppose that intrasite oscillations $\tilde{u}_n$ about
the centre mass are small and can be considered to be harmonical;
believe that the probability of charge's occurence
on sites depends linearly on sites displacements $\tilde{u}_n$,
$\alpha'$ is a coupling constant,
$M$ is the $n$th site's effective mass, 
$K$ is the elastic constant.
Motion equations of Hamiltonian~\eqref{eq1} have the form:
\begin{align}
i \hbar \frac{d b_n}{d \tilde{t}} &=
\nu_{n,n-1} b_{n-1} + \nu_{n,n} b_n  
+
\nu_{n,n+1} b_{n+1} + \alpha' \tilde{u}_n b_n,
\label{eq2} \\
M \frac{d^2 \tilde{u}_n}{d \tilde{t}^2} &=
-K \tilde{u}_n - \alpha' |b_n|^2 -
\tilde{\gamma} \frac{d \tilde{u}_n}{d \tilde{t}}
+ \tilde{A}_n (\tilde{t}).
\label{eq3}
\end{align}
To model a thermostat, subsystem \eqref{eq3} involves the term with friction
($\tilde{\gamma}$ is the friction coefficient) and the random
force $\tilde{A}_n (\tilde{t})$ such that
$\langle \tilde{A}_n (\tilde{t}) \rangle =0$,
$\langle \tilde{A}_n (\tilde{t}) \tilde{A}_m (\tilde{t}+\tilde{s}) \rangle =
2 k_B T \tilde{\gamma}_n \delta_{nm} \delta (\tilde{s})$
($T$ is the temperature [K], $k_B$ -- Boltzmann constant).
This way of imitating the environmental
temperature with the use of Langevin equations \eqref{eq3}
is well known \cite{bib7,bib8,bib81}.

\section{On the calculation of the diffusion coefficient}
\label{SecParam}

We assessed the charge mobility in the following way \cite{bib9,bib10}.
To calculate the mobility $\mu$, one should find the time dependence of the
mean-root-square displacement averaged over simulations
$\langle X^2(t) \rangle = \langle \sum_{n=1}^N |b_n|^2 n^2 \rangle$
at a given temperature $T$ and then use it to derive the diffusion
coefficient $D$ which enables one to assess the charge mobility $\mu$ in the
chain. Individual simulations are trajectories of the ordinary
differential equations system from various initial conditions and with
various values of the random force simulating the thermostat.

To nondimensionlize system \eqref{eq2},\eqref{eq3}
let us choose arbitrary characteristic time $\tau$,
$\tilde{t}=\tau t$,
and a characteristic size of displacement $U^*_n$, $\tilde{u}_n =U^*_n u_n$.
For a homogeneous chain, the nondimensionalized
motion equations, determining the distribution of the charge along
$N$-site chain, have the form:
\begin{align}
i \frac{d b_n}{d t} &=
\eta ( b_{n-1} + b_{n+1} ) + \chi u_n b_n,
\label{eq2nondim} \\
\frac{d^2 u_n}{d t^2} &=
-\omega^2 u_n - \chi |b_n|^2 - \gamma \frac{d u_n}{d t}
+ \xi Z_n (t).
\label{eq3nondim}
\end{align}
The relations between dimension and dimensionless parameters are as
follows. Matrix elements $\eta = \nu_{n,n\pm1} \tau / \hbar$,
frequencies of sites oscillation $\omega = \tau \sqrt{K/M}$,
$\gamma = \tau \tilde{\gamma} / M$.
The characteristic size of displacements $U^* = \sqrt{\hbar \tau / M}$
is chosen such that the multiplier of the terms
in \eqref{eq2nondim} and \eqref{eq3nondim} which are responsible for
the interaction between the quantum and classical subsystems
be the same, the coupling constant
$\chi = \alpha' \sqrt{ \tau^3 / \hbar M}$.
$Z_n(t)$ is a Gaussian random variable with the distribution
\begin{align}
\langle Z_n(t) \rangle = 0, \quad
\langle Z_n(t)  Z_n(t+t') \rangle = \delta(t'),
\nonumber \\
\xi = \frac{\sqrt{ 2 k_B T \tilde{\gamma} \tau^3} }{M U^*} =
\sqrt{ \frac{2 k_B T^* \tau}{\hbar}} \sqrt{ \gamma \mathrm{T}},
\label{temperxi}
\end{align}
where the dimensionless temperature is $\mathrm{T} = T/T^*$.
In modeling we believe that the parameters of classical sites are
the same, and the value of the matrix element $\eta$ depends on the
nucleotide sequence type.

The parameters of the model corresponding to the DNA fragment are the
following: the characteristic time is $\tau = 10^{-14}$\,s
(we chose time scale corresponding to quantum subsystem \eqref{eq2nondim}),
the effective mass of a complementary pair is $M = 10^{-21}$\,g.
The dimensionless coefficients are:
frequencies of classical sites $\omega = 0.01$ (which corresponds to
the spring rigidity
$K \approx 0.06 $\,eV/\AA$^2$
of hydrogen bonds between complementary
bases), $\chi = 0.02$ ($\alpha' \approx 0.13\,$eV/\AA),
friction coefficient is $\gamma = 0.006$ ($\tilde{\gamma}/M =
6\cdot10^{11}\,\text{s}^{-1}$),
for the chosen characteristic temperature $T^* = 1$\,K,
coefficient $\xi \approx 0.051 \sqrt{\gamma \mathrm{T}}$.
The values of the matrix elements which were used in calculations
\cite{bib11,bib12} are given in Table \ref{tabl1}.

\begin{table}[hbt]
\caption{Dimension and dimensionless values of matrix elements of the
transition between sites \cite{bib11,bib12}.
\label{tabl1}}
\begin{center}
\begin{tabular}{ccc}
\\ \hline
sequence type & $\nu$, eV & $\eta$ \\
\hline
polyA & 0.030 & 0.456 \\
polyC & 0.041 & 0.623 \\
polyG & 0.084 & 1.276 \\
polyT & 0.158 & 2.400 \\
\hline
\end{tabular}
\end{center}
\end{table}

Integrating numerically system \eqref{eq2nondim}, \eqref{eq3nondim}
from given initial conditions (at $t=0$ classical displacements and site
velocities are determined from the thermodynamic equilibrium
distribution, and the charge is considered to
be localized on one site in the center of the chain)
we find the charge dynamics and sites' trajectories
at a given temperature in an individual simulation.
Then we calculate $\langle X^2(t)\rangle$ averaged over
simulations and use it to find the diffusion coefficient $\mathrm{D}$
at a given ``temperature'' $\mathrm{T}$:
\begin{align}
\langle X^2(t)\rangle = \bigg\langle \sum_{n= -N/2}^{N/2}
|b_n(t)|^2 n^2 \bigg\rangle, \quad \langle X^2(t)\rangle = 2 \mathrm{D} t.
\label{eqX2}
\end{align}
Calculations of individual simulations
were carried out by 2o2s1g-method \cite{bib13}.
The model parameters as applied to DNA are given
in more detail, for example, in \cite{bib15}.

\section{Main results}
\label{SecScale}

Since we are interested in the qualitative picture, all the results
are presented in dimensionless form. A change to dimension
values is simple. Here we give a formula to assess the
dimension value of the mobility $\mu(T)$ at a given temperature $T$ [K]
from the calculated dimensionless value of the diffusion coefficient
$\mathrm{D}(\mathrm{T})$:
\begin{align}
\mu = \frac{\mathrm{D}}{\mathrm{T}} \frac{e \tau a^2}{k_B T^*},
\end{align}
where $a$ is the distance between neighboring sites of the molecular
chain, $e$ is the electron charge,
$\mathrm{T} = T/ T^*$. For DNA $a \approx 3.4$\,\AA.

\begin{figure}[htb]
\includegraphics[width=0.42\textwidth]{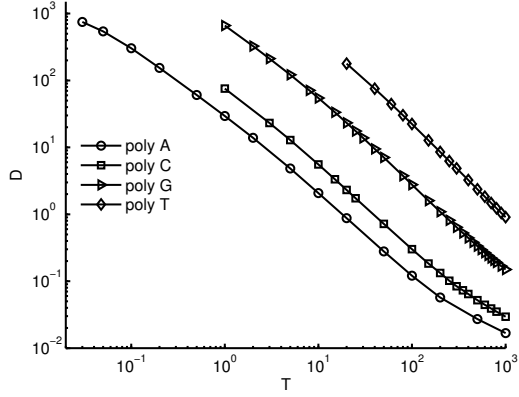}
\caption{Dependencies of the diffusion coefficient $\mathrm{D}$ on temperature $\mathrm{T}$
for homogeneous nucleotides.
Calculation results denoted by symbols are connected by line segments.
The scales are logarithmic.
\label{fig1}}
\end{figure}

For different sequence types occurring at the same
temperature $\mathrm{T}$, clearly,
the calculated values of $\mathrm{D}$ are different
(the difference between polyA and polyT is nearly two orders
of magnitude, see Fig.\ \ref{fig1}). However the
temperature dependence of the diffusion $\mathrm{D}(\mathrm{T})$
seems to be alike in all the cases.

\begin{figure}[htb]
\includegraphics[width=0.48\textwidth]{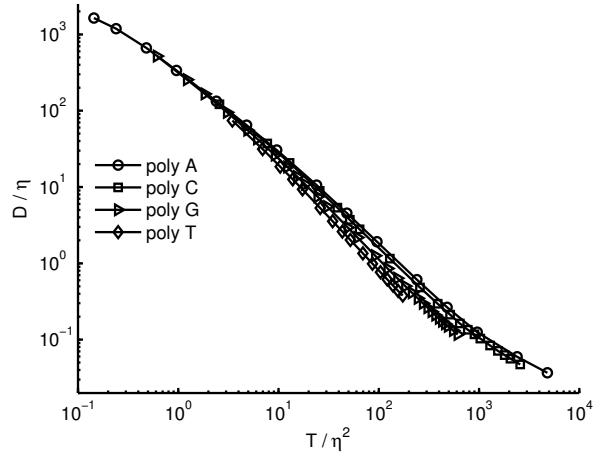}
\caption{Rescaled 
dependencies for homogeneous polynucleotides.
\label{fig2}}
\end{figure}

It turned out that upon rescaling
$\mathrm{T} \to \mathrm{T}/\eta^2$, 
$\mathrm{D} \to \mathrm{D}/\eta$, 
all the graphs are similar, especially for low temperatures (see Fig.\ \ref{fig2}).

\begin{figure}[htb]
\includegraphics[width=0.45\textwidth]{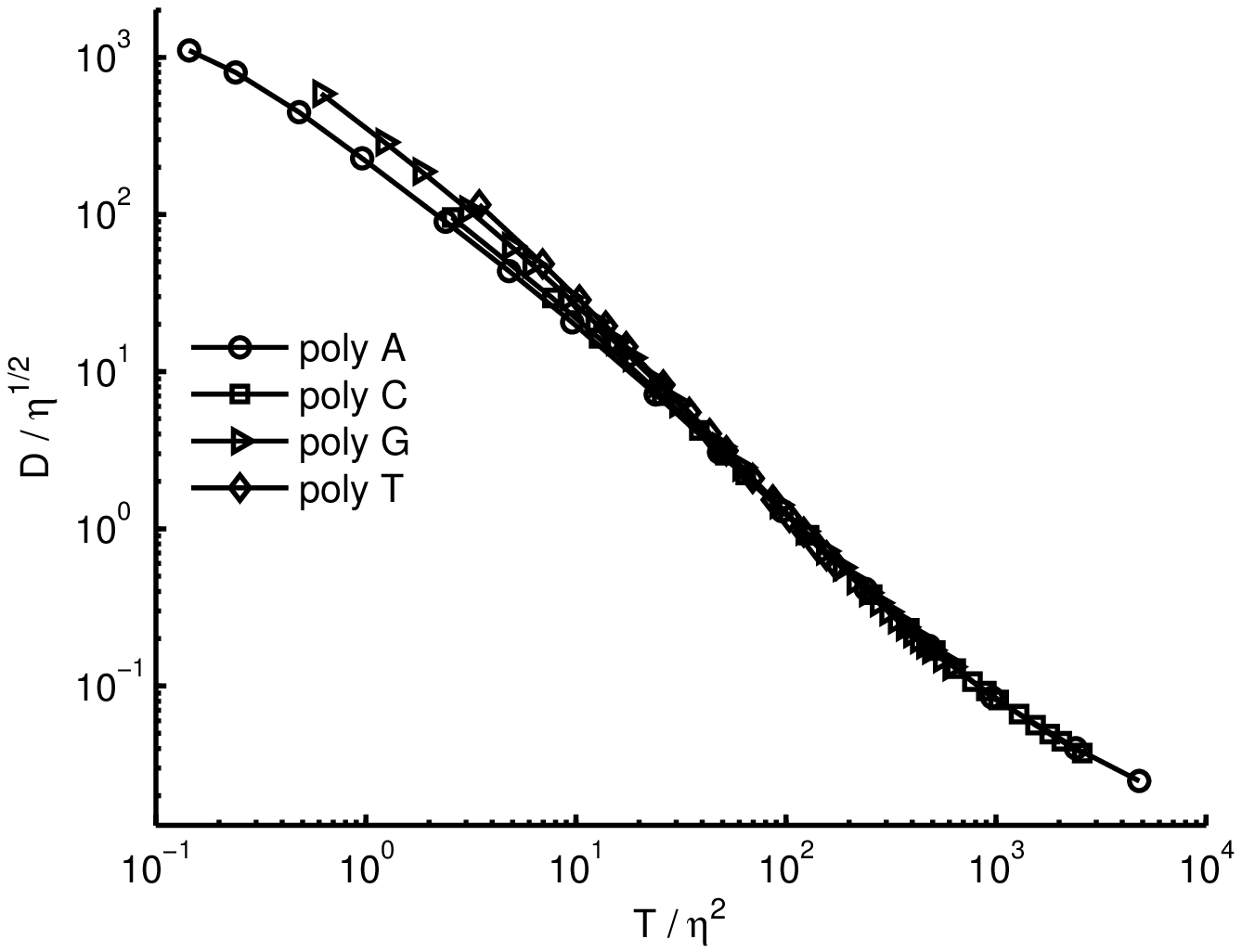}
\caption{Other rescaled 
dependencies for homogeneous nucleotides.
The temperature $\mathrm{T}$ is rescaled  as in Fig.\ \ref{fig1},
but the diffusion coefficient $\mathrm{D}$ is divided by $\sqrt{\eta}$.
\label{fig3}}
\end{figure}

It has been empirically found that for medium temperatures,
rescaling $\mathrm{T} \to \mathrm{T}/\eta^2 $, 
$\mathrm{D} \to \mathrm{D}/ \sqrt{\eta} $ 
suits better. From Fig.\ \ref{fig3} we
notice that for rescaled temperature $\mathrm{T}/\eta^2  > 10$,
all the graphs are very close to one another.

\begin{figure}
\includegraphics[width=0.48\textwidth]{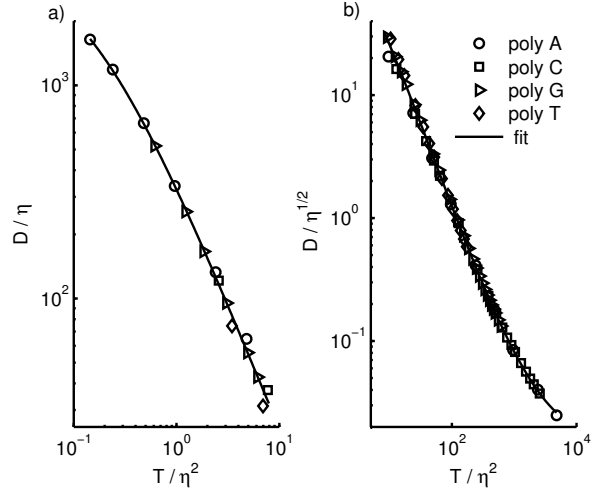}
\caption{Results of modeling and graphs of approximating polynomials.
a) Low temperatures, rescaling
$\mathrm{T} \to \mathrm{T}/\eta^2 $, $\mathrm{D} \to \mathrm{D}/\eta $,
b) medium temperatures, rescaling
$\mathrm{T} \to \mathrm{T}/\eta^2$,
$\mathrm{D} \to \mathrm{D}/\sqrt{\eta}$.
\label{fig4}}
\end{figure}

We approximated the data on a logarithmic scale by cubic polynomial
for both the dependencies $\mathrm{D}_1 = \mathrm{D}/\eta$
and $\mathrm{D}_2 = \mathrm{D}/\sqrt{\eta}$ on different
temperature intervals $\mathrm{T}_1 = \mathrm{T}/\eta^2$, having chosen
$0 < \mathrm{T}_1 \leq R \approx 8$ for the
approximation interval $\mathrm{D}_1$,
and $\mathrm{T}_1 > R$ for $\mathrm{D}_2$.
The obtained parameter values of the functions
\begin{align}
y=a_0 x^3 + a_1 x^2 + a_2 x + a_3,\quad x = \ln(\mathrm{T}/ \eta^2),
\label{appr2}
\end{align}
are as follows:
\begin{itemize}
\item[(I)]
for $y_1 = \ln(\mathrm{D}/ \eta)$, on the interval $0<\mathrm{T}/ \eta^2 \leq 8$,
\begin{align*}
a_0 &= 1.3359590\cdot10^{-2}, & a_1 &= -7.0449850\cdot10^{-2}, \\
a_2 &= -1.0275530, & a_3 &= 5.7815836 ;
\end{align*}
\item[(II)]
for $y_2 = \ln(\mathrm{D}/\sqrt{ \eta})$, on the interval $8 \leq \mathrm{T}/ \eta^2$,
\begin{align*}
a_0 &= 1.4621272\cdot10^{-2} , & a_1 &= -1.7419911\cdot10^{-1} , \\
a_2 &= -6.5194332\cdot10^{-1} , & a_3 &= 5.4939738 .
\end{align*}
\end{itemize}
Graphs of approximating polynomials are shown in Fig.~\ref{fig4}.

The boundary value $R \approx 8$ is chosen in the following way. For the
results from the intercept $\mathrm{T}_1 \in [0,r]$, we calculated
a summary deviation $S_1$
from the approximation curve $y_1(x)$, normalized by the number of
results $K$, and on the second interval $r < \mathrm{T}_1$
we found the same deviation $S_2$ from $y_2(x)$:
\begin{align*}
S_1 &= \frac{1}{K} \sum_{k=1}^{K} [y_1(x_k) - y_{1k}],
\quad S_2 = \frac{1}{L} \sum_{l=1}^{L} [y_2(x_l) - y_{2l}].
\end{align*}
With increasing $r$, $S_1$ grows and $S_2$ decreases.
The abscissa of intersection of their graphs is chosen
to be a boundary of partitions $R$.

\section{Extensions of the model. Taking account of stacking interaction in DNA}
\label{SecAdd}

In DNA, of great
importance is nonlinear stacking interaction $\Theta(\tilde{u}_n - \tilde{u}_{n-1})$
\cite{bib16,bib162},
which for small values of the difference has the form \cite{bib16}
\begin{align*}
\Theta(\tilde{u}_n - \tilde{u}_{n-1}) = \frac 12 K_s \cdot (\tilde{u}_n - \tilde{u}_{n-1})^2.
\end{align*}
More detailed Peyrard--Bishop model with nonlinear interaction between
neighboring base pairs \cite{bib161,bib162} for the case of small site displacements
can be reduced to the form similar to that of the dispersion term in equations for crystals.

Solvation effects play a great role in processes of charge transfer \cite{bib17,bib18}.

Earlier \cite{bib15} we have calculated the hole mobility for polyG fragments of DNA
in the cases of
dispersion in classical chain and taking account of solvation effects.
The total energy of the system has the form
\begin{align*}
\langle \Psi | \hat{H} |\Psi \rangle =
\sum_{m,n} \nu_{nm} b_m b_{n}^*
+
\sum_n \alpha' \tilde{u}_{n}  b_{n} b_{n}^* \\
+
\frac 12 \sum_n \tilde{\Phi} (b_{n} b_{n}^*)^2
+
\frac 12 \sum_n K_s (\tilde{u}_n - \tilde{u}_{n-1})^2 \\
+
\frac 12 \sum_n M \dot{\tilde{u}}_{n}^2
+
\frac 12 \sum_n K \tilde{u}_{n}^2
.
\end{align*}
Here $K_s$ is a constant determining the contribution of dispersion into the chain energy.
In molecular crystals, the value of dispersion $K_s$  in a classical chain is usually small. 
For DNA, this is not the case.
For DNA in \cite{bib16}
the value of stacking interaction was found to be $K_s \approx 0.04$\,eV/AA$^2$,
and for intramolecular 
hydrogen bonds $K \approx 0.06$ eV/AA$^2$.

The energy of charge's solvation on the $n$-th site depends on the charge distribution density on the site \cite{ref3-1},
$\tilde{\Phi}$ is the effective solvation coefficient. In the calculations of diffusion coefficient we took
$\tilde{\Phi} = 1.04$\,eV \cite{bib18}.

The dimensionless parameters are: $k_{s} = 6.4 \cdot 10^{-5}$, $\Phi = 15.5$.
In calculations of individual simulations we added random force and friction into motion equations of classical sites, as aforesaid.

Using an approximating curve obtained for medium temperatures (II),
we considered an ``inverse problem'' as applied to
temperature dependencies D(T), founded for model with dispersion and solvation.

I.e., for homogeneous chains we have found the cubic polynomial approximation \eqref{appr2} with coefficients $a_i$ from (II)
in the coordinate system $x = \ln(\mathrm{T}/ \eta^2)$, $y = \ln(\mathrm{D}/ \sqrt{\eta})$.
Let us assume that for a certain chain (with dispersion or with solvation, or a regular chain),
we can find an ``effective'' value of $\eta_{\mathit{eff}}$ such that upon rescaling, the graph
D(T) for this chain  will fall on this approximating curve (II).

This problem reduces to finding a minimum of the distance $R(\eta )$ from a point to the curve (II).
We have data for one temperature (T,D). It is required to find $\eta$, such that the point with the coordinates
$x_0 = \ln(\mathrm{T}/ \eta^2), y0 = \ln(\mathrm{D}/ \sqrt{\eta})$  be as close to curve (II) as possible.
If $\eta_{\mathit{eff}}$ values obtained are close for different values T, then the graph will be similar
to the graph of D(T) in a homogeneous chain with the matrix element $\eta_{\mathit{eff}}$.

The test of this assumption showed that in the parameter range under consideration:
1)~For chains with dispersion ($k_s = 6.4\cdot 10^{-5}$, $\Phi = 0$) it is valid;
2)~For chains with solvation, chains with solvation and dispersion and regular chains it is not valid.

\begin{figure}
\includegraphics[width=0.48\textwidth]{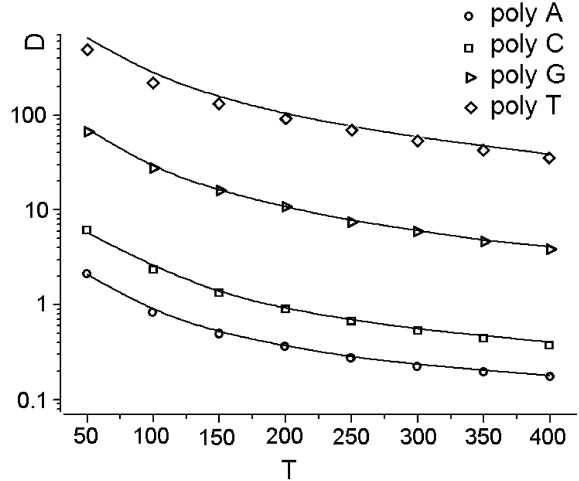}
\caption{Dimensionless temperature dependencies of the mobility, ${\mathrm T}T^*=T$\,[K],
in semilogarithmic scale. Symbols stand for the values of the diffusion coefficient D,
calculated for homogeneous chains with dispersion.
Continuous lines going adjacently join the values of D(T) calculated for dispersionless chains
(with the use of equations \eqref{eq2nondim}, \eqref{eq3nondim}) with other matrix elements:
near the values for polyA runs a curve with $\eta_{\mathit{eff}}  = 0.7$,
near those for polyC is a curve with  $\eta_{\mathit{eff}} = 0.96$,
near the values for polyG -- a curve with  $\eta_{\mathit{eff}} = 2.08$,
and near those for polyT -- a curve with  $\eta_{\mathit{eff}} = 4.1$.
\label{fighom}}
\end{figure}

For the first case, we calculated diffusion coefficient D(T) in all the homogeneous polynucleotide chains with dispersion.
The $\eta_{\mathit{eff}}$ values were close for different temperatures. The D(T) in polyA fragment was found
to be close to that in a homogeneous chain with $\eta_{\mathit{eff}} \approx 0.70$
(which corresponds to the matrix element $\nu  \approx 0.046$\,eV),
for polyC  $\eta_{\mathit{eff}} \approx 0.96$ ($\nu \approx 0.063$\,eV);
for polyG $\eta_{\mathit{eff}} \approx 2.08$ ($\nu \approx 0.137$\,eV),
and for polyT  $\eta_{\mathit{eff}} \approx 4.1$ ($\nu  \approx 0.270$\,eV).
The results of the calculations in chains with dispersion and in homogeneous chains with ``effective''
matrix elements $\eta_{\mathit{eff}}$ are shown in Figure \ref{fighom}.
A considerable discrepancy for polyT at $\mathrm{T} \le 150$ stems from the fact that the ``boundary value'' $R \approx 8$,
below which another approximation (I) should be used, for $\eta = 4.1$ corresponds to $\mathrm{T} = R \eta^2 \approx 135$.

So, to find temperature dependence $\mathrm{D}(\mathrm{T})$ of the charge in the homogeneous chain with dispersion
we may to calculate $\mathrm{D}$ for one value $\mathrm{T}$ and to count $\eta_{\mathit{eff}}$.
Than, this $\eta_{\mathit{eff}}$ may be used for estimating $\mathrm{D}$ at different temperatures.

For homogeneous chains with solvation $\Phi = 15.5$, we failed to find a common matrix element $\eta_{\mathit{eff}}$
for different temperatures (see Table \ref{tabnu}).

We also calculated mobility for regular fragments of the form of {\ldots}ATATAT{\ldots} and {\ldots}GTGTGT{\ldots}.
In calculations of individual simulations, integration was performed for the system of equations \eqref{eq2nondim}, \eqref{eq3nondim},
in which matrix elements depended on the sequence type and the sites of the classical subsystem were assumed to be similar.
The values of matrix elements 
were taken from \cite{bib11,bib12}:
$\nu_{\mathrm{AT}} = 0.105$\,eV ($\eta_{\mathrm{AT}} = 1.595$), $\nu_{\mathrm{TA}} = 0.086$\,eV ($\eta_{\mathrm{TA}}  = 1.307$),
$\nu_{\mathrm{GT}} = 0.137$\,eV ($\eta_{\mathrm{GT}}  = 2.081$), $\nu_{\mathrm{TG}} = 0.085$\,eV ($\eta_{\mathrm{TG}} = 1.291$).
Then we tried to fit $\eta_{\mathit{eff}}$, however $\eta_{\mathit{eff}}$ differs considerably
for different temperatures (see Table~\ref{tabnu}).

\begin{table}[htb]
\caption{Results of $\eta_{\mathit{eff}}$ calculation for the minimum distance to the approximating curve (II).
Value of the effective matrix element $\eta_{\mathit{eff}}$ for different temperature T.
\label{tabnu}}
\begin{center}
\begin{tabular}{cccc}
\\ \hline
Chain & T=100 & T=200 & T=300 \\
\hline
polyG with solvation & & & \\
($\Phi = 15.5$, $k_s = 0$)
& 0.33 & 0.28 & 0.22 \\
\hline
polyG with dispersion and solvation & & & \\
($\Phi = 15.5$, $k_s = 6.4\cdot 10^{-5}$)
&0.62 & 0.57 & 0.53 \\
\hline
{\ldots}ATATAT{\ldots} &
0.77 & 0.82 & 0.88\\
\hline
{\ldots}GTGTGT{\ldots} &
0.47 & 0.52 & 0.55  \\
\hline
\end{tabular}
\end{center}
\end{table}

As can be seen from Table \ref{tabnu}, for the case of chain with solvation we could not find
a single value $\eta_{\mathit{eff}}$ for different temperatures.
Also, the idea of $\eta_{\mathit{eff}}$ is not worked for regular polynucleotides.

\section{Discussion}
\label{SecEnd}

It is shown that the values of the diffusion coefficient $\mathrm{D}$ in a
Holstein model chain simulating homogeneous DNA,
which are different for different nucleotide types,
being rescaled, fall on one and the same curve in the corresponding
range $\mathrm{T}_1 = \mathrm{T}/ \eta^2$.
For low $\mathrm{T}_1 \leq 8$, rescaling is
$\mathrm{D} \to \mathrm{D} / \eta$,
for medium $T_1 > 8$, rescaling $\mathrm{D} \to \mathrm{D} / \sqrt{\eta}$
suits better. For these data on a logarithmic scale,
approximating cubic dependencies are found.

In studying the charge mobility in system \eqref{eq2nondim}, \eqref{eq3nondim},
we were interested in a qualitative picture and left aside the domain
of applicability of the model.
The semiclassical model used cannot be applied at temperatures
below Debye one $k_B T \leq \Theta = \hbar \omega$
(for model nucleotide pairs $\Theta \approx 8$\,K).
Calculations were carried out for coefficients that are similar
at any temperature.
This is a simplest assumption. Surely some parameters are temperature
dependent. The most spectacular example is concerned with DNA whose
constant of hydrogen bonds interaction $K \to 0$ as $T \to 350$\,K
(at temperature 60--80$^\circ$C DNA melts
and hydrogen bonds of complementary pairs are broken).
It may be assumed that as the temperature decreases, the coefficient
values change less and less and finally become a constant.

We considered the Holstein
model of DNA where Watson--Crick pairs are represented as independent oscillators
described by classical motion equations.
It is believed that the planes of nucleotide base
pairs are parallel to each other at any moment and the distances between
neighboring planes are unchanged (the standard DNA model). The transfer of a hole in a DNA
is determined by overlapping of its wave functions at neighboring sites.
In view of the model geometry,
the overlapping integrals are virtually independent of the displacements.
Thus in Hamiltonian \eqref{eq1} we take into account the
(intrasite) displacements for the diagonal matrix elements only.

In Su-Schrieffer-Heeger (SSH) model \cite{refer4}, the non-diagonal matrix element dependence
on inter-site displacements is considered.
SSH model has been applied to DNA by the Conwell et al.\ \cite{refer5,refer6,refer7}.
Two important degrees of freedom in DNA chain
are relative base pair displacements
along the stack and the relative twist angles.
It was shown \cite{refer6} that since these degrees of freedom
are not independent they can be taken into account by
introducing the dependence of the matrix elements on
the inter-site displacements with effective coupling constant.
SSH model was applied to describe the properties of polarons in DNA
in many works (see e.g.\ \cite{refer71,refer8,refer9} and references therein).
In the work \cite{refer10} we calculated the hole mobility for
Holstein model, SSH model and combined one (HSSH-model), in polyG at $T=300$\,K.
The values obtained were similar.
It is task for further research to verify the scaling laws for SSH and HSSH DNA model.

We considered the simple case of the harmonic potentials in the classical chain of sites.
Holstein model with dispersion exactly corresponds to the Peyrard--Bishop model for DNA\cite{bib161},
when sites displacements from their equilibrium positions are small\cite{bib162}.
This should undoubtedly be valid for low temperatures, however at room and higher temperatures
the assumption of the displacements smallness can be incorrect.
In this case
consideration of the Peyrard--Bishop model which takes account of the chain anharmonicity becomes actual.
The authors are planning to study this problem in the future.

Based on the numerical results for Holstein semiclassical model,
we can assume that charge mobility in molecular chain with dispersion and matrix element $\eta_1$
looks like mobility in chain without dispersion and matrix element $\eta_2$, and $\eta_1 < \eta_2$.
Also, the approximated cubic curve is not valid for regular chains and for homogeneous chain with solvation.
The curve can be applied for estimation of the hole mobility in ``dry DNA'' rather than in ``DNA in a solvent''.

We believe that in the range of ``biologically significant''
temperatures, for homogeneous biopolymers with parameters close to DNA
parameters discussed, one can approximately assess the value of the
charge mobility at temperature $T$ without carrying out
resource-intensive model calculations, just by associating it with a
point with suitable coordinate $\mathrm{T}_1$
and recalculating the diffusion coefficient.

\begin{acknowledgments}
%
We would like to thank the referees for their careful reading as well as many helpful comments,
which have led to improvements of the paper.

We are grateful for providing us with the computational resources.
Calculations were made in the Joint Supercomputer Center RAS and
Supercomputing Center of Lomonosov Moscow State University.
The reported study was partially supported by
Russian Foundation for Basic Research, research projects
No.\ 14-07-00894, 13-07-00256, 13-07-00331, 12-07-00279,
12-07-33006-mol-a-ved.
\end{acknowledgments}

\bibliography{fialka_rev3}

\begin{thebibliography}{33}%
\makeatletter
\providecommand \@ifxundefined [1]{%
 \@ifx{#1\undefined}
}%
\providecommand \@ifnum [1]{%
 \ifnum #1\expandafter \@firstoftwo
 \else \expandafter \@secondoftwo
 \fi
}%
\providecommand \@ifx [1]{%
 \ifx #1\expandafter \@firstoftwo
 \else \expandafter \@secondoftwo
 \fi
}%
\providecommand \natexlab [1]{#1}%
\providecommand \enquote  [1]{``#1''}%
\providecommand \bibnamefont  [1]{#1}%
\providecommand \bibfnamefont [1]{#1}%
\providecommand \citenamefont [1]{#1}%
\providecommand \href@noop [0]{\@secondoftwo}%
\providecommand \href [0]{\begingroup \@sanitize@url \@href}%
\providecommand \@href[1]{\@@startlink{#1}\@@href}%
\providecommand \@@href[1]{\endgroup#1\@@endlink}%
\providecommand \@sanitize@url [0]{\catcode `\\12\catcode `\$12\catcode
  `\&12\catcode `\#12\catcode `\^12\catcode `\_12\catcode `\%12\relax}%
\providecommand \@@startlink[1]{}%
\providecommand \@@endlink[0]{}%
\providecommand \url  [0]{\begingroup\@sanitize@url \@url }%
\providecommand \@url [1]{\endgroup\@href {#1}{\urlprefix }}%
\providecommand \urlprefix  [0]{URL }%
\providecommand \Eprint [0]{\href }%
\providecommand \doibase [0]{http://dx.doi.org/}%
\providecommand \selectlanguage [0]{\@gobble}%
\providecommand \bibinfo  [0]{\@secondoftwo}%
\providecommand \bibfield  [0]{\@secondoftwo}%
\providecommand \translation [1]{[#1]}%
\providecommand \BibitemOpen [0]{}%
\providecommand \bibitemStop [0]{}%
\providecommand \bibitemNoStop [0]{.\EOS\space}%
\providecommand \EOS [0]{\spacefactor3000\relax}%
\providecommand \BibitemShut  [1]{\csname bibitem#1\endcsname}%
\let\auto@bib@innerbib\@empty
\bibitem [{\citenamefont {Lakhno}(2008)}]{bib1}%
  \BibitemOpen
  \bibfield  {author} {\bibinfo {author} {\bibfnamefont {V.}~\bibnamefont
  {Lakhno}},\ }\href@noop {} {\bibfield  {journal} {\bibinfo  {journal}
  {International Journal of Quantum Chemistry}\ }\textbf {\bibinfo {volume}
  {108}},\ \bibinfo {pages} {1970} (\bibinfo {year} {2008})}\BibitemShut
  {NoStop}%
\bibitem [{\citenamefont {Offenh\"ausser}\ and\ \citenamefont
  {Rinaldi}(2009)}]{bib2}%
  \BibitemOpen
  \bibinfo {editor} {\bibfnamefont {A.}~\bibnamefont {Offenh\"ausser}}\ and\
  \bibinfo {editor} {\bibfnamefont {R.}~\bibnamefont {Rinaldi}},\ eds.,\
  \href@noop {} {\emph {\bibinfo {title} {Nanobioelectronics -� for
  {E}lectronics, {B}iology, and {M}edicine.}}}\ (\bibinfo  {publisher}
  {Springer, New York},\ \bibinfo {year} {2009})\ p.\ \bibinfo {pages}
  {337}\BibitemShut {NoStop}%
\bibitem [{\citenamefont {Alexandre}\ \emph {et~al.}(2003)\citenamefont
  {Alexandre}, \citenamefont {Artacho}, \citenamefont {Soler},\ and\
  \citenamefont {Chacham}}]{refer2-1}%
  \BibitemOpen
  \bibfield  {author} {\bibinfo {author} {\bibfnamefont {S.~S.}\ \bibnamefont
  {Alexandre}}, \bibinfo {author} {\bibfnamefont {E.}~\bibnamefont {Artacho}},
  \bibinfo {author} {\bibfnamefont {J.}~\bibnamefont {Soler}}, \ and\ \bibinfo
  {author} {\bibfnamefont {H.}~\bibnamefont {Chacham}},\ }\href@noop {}
  {\bibfield  {journal} {\bibinfo  {journal} {Physical Review Letters}\
  }\textbf {\bibinfo {volume} {91}},\ \bibinfo {pages} {108105} (\bibinfo
  {year} {2003})}\BibitemShut {NoStop}%
\bibitem [{\citenamefont {Wang}\ \emph {et~al.}(2006)\citenamefont {Wang},
  \citenamefont {Fu},\ and\ \citenamefont {Wang}}]{refer2-2}%
  \BibitemOpen
  \bibfield  {author} {\bibinfo {author} {\bibfnamefont {Y.}~\bibnamefont
  {Wang}}, \bibinfo {author} {\bibfnamefont {L.}~\bibnamefont {Fu}}, \ and\
  \bibinfo {author} {\bibfnamefont {K.-L.}\ \bibnamefont {Wang}},\ }\href@noop
  {} {\bibfield  {journal} {\bibinfo  {journal} {Biophysical Chemistry}\
  }\textbf {\bibinfo {volume} {119}},\ \bibinfo {pages} {107} (\bibinfo {year}
  {2006})}\BibitemShut {NoStop}%
\bibitem [{\citenamefont {Starikov}(2005)}]{refer2-3}%
  \BibitemOpen
  \bibfield  {author} {\bibinfo {author} {\bibfnamefont {E.}~\bibnamefont
  {Starikov}},\ }\href@noop {} {\bibfield  {journal} {\bibinfo  {journal}
  {Philosophical Magazine}\ }\textbf {\bibinfo {volume} {85}},\ \bibinfo
  {pages} {3435} (\bibinfo {year} {2005})}\BibitemShut {NoStop}%
\bibitem [{\citenamefont {Fialko}\ and\ \citenamefont
  {Lakhno}(2000)}]{refer2-4}%
  \BibitemOpen
  \bibfield  {author} {\bibinfo {author} {\bibfnamefont {N.}~\bibnamefont
  {Fialko}}\ and\ \bibinfo {author} {\bibfnamefont {V.}~\bibnamefont
  {Lakhno}},\ }\href@noop {} {\bibfield  {journal} {\bibinfo  {journal}
  {Physical Letters A}\ }\textbf {\bibinfo {volume} {278}} (\bibinfo {year}
  {2000})}\BibitemShut {NoStop}%
\bibitem [{\citenamefont {Dauxois}\ \emph {et~al.}(1993)\citenamefont
  {Dauxois}, \citenamefont {Peyrard},\ and\ \citenamefont {Bishop}}]{bib162}%
  \BibitemOpen
  \bibfield  {author} {\bibinfo {author} {\bibfnamefont {T.}~\bibnamefont
  {Dauxois}}, \bibinfo {author} {\bibfnamefont {M.}~\bibnamefont {Peyrard}}, \
  and\ \bibinfo {author} {\bibfnamefont {A.}~\bibnamefont {Bishop}},\
  }\href@noop {} {\bibfield  {journal} {\bibinfo  {journal} {Physical Review
  E}\ }\textbf {\bibinfo {volume} {47}},\ \bibinfo {pages} {R44} (\bibinfo
  {year} {1993})}\BibitemShut {NoStop}%
\bibitem [{\citenamefont {Holstein}(1959)}]{bib3}%
  \BibitemOpen
  \bibfield  {author} {\bibinfo {author} {\bibfnamefont {T.}~\bibnamefont
  {Holstein}},\ }\href@noop {} {\bibfield  {journal} {\bibinfo  {journal}
  {Annals of Physics}\ }\textbf {\bibinfo {volume} {8}},\ \bibinfo {pages}
  {325} (\bibinfo {year} {1959})}\BibitemShut {NoStop}%
\bibitem [{\citenamefont {Henderson}\ \emph {et~al.}(1999)\citenamefont
  {Henderson}, \citenamefont {Jones}, \citenamefont {Hampikian}, \citenamefont
  {Kan},\ and\ \citenamefont {Schuster}}]{bib4}%
  \BibitemOpen
  \bibfield  {author} {\bibinfo {author} {\bibfnamefont {P.}~\bibnamefont
  {Henderson}}, \bibinfo {author} {\bibfnamefont {D.}~\bibnamefont {Jones}},
  \bibinfo {author} {\bibfnamefont {G.}~\bibnamefont {Hampikian}}, \bibinfo
  {author} {\bibfnamefont {Y.}~\bibnamefont {Kan}}, \ and\ \bibinfo {author}
  {\bibfnamefont {G.}~\bibnamefont {Schuster}},\ }\href@noop {} {\bibfield
  {journal} {\bibinfo  {journal} {PNAS USA}\ }\textbf {\bibinfo {volume}
  {96}},\ \bibinfo {pages} {8353} (\bibinfo {year} {1999})}\BibitemShut
  {NoStop}%
\bibitem [{\citenamefont {Grozema}\ \emph {et~al.}(2000)\citenamefont
  {Grozema}, \citenamefont {Berlin},\ and\ \citenamefont {Siebbeles}}]{bib5}%
  \BibitemOpen
  \bibfield  {author} {\bibinfo {author} {\bibfnamefont {F.}~\bibnamefont
  {Grozema}}, \bibinfo {author} {\bibfnamefont {Y.}~\bibnamefont {Berlin}}, \
  and\ \bibinfo {author} {\bibfnamefont {L.}~\bibnamefont {Siebbeles}},\
  }\href@noop {} {\bibfield  {journal} {\bibinfo  {journal} {Journal of the
  American Chemical Society}\ }\textbf {\bibinfo {volume} {122}},\ \bibinfo
  {pages} {10903} (\bibinfo {year} {2000})}\BibitemShut {NoStop}%
\bibitem [{\citenamefont {Fialko}\ and\ \citenamefont {Lakhno}(2002)}]{bib6}%
  \BibitemOpen
  \bibfield  {author} {\bibinfo {author} {\bibfnamefont {N.}~\bibnamefont
  {Fialko}}\ and\ \bibinfo {author} {\bibfnamefont {V.}~\bibnamefont
  {Lakhno}},\ }\href@noop {} {\bibfield  {journal} {\bibinfo  {journal}
  {Regular \& Chaotic Dynamics}\ }\textbf {\bibinfo {volume} {7}},\ \bibinfo
  {pages} {299} (\bibinfo {year} {2002})}\BibitemShut {NoStop}%
\bibitem [{\citenamefont {Turg}\ \emph {et~al.}(1977)\citenamefont {Turg},
  \citenamefont {Lantelme},\ and\ \citenamefont {Friedman}}]{bib7}%
  \BibitemOpen
  \bibfield  {author} {\bibinfo {author} {\bibfnamefont {P.}~\bibnamefont
  {Turg}}, \bibinfo {author} {\bibfnamefont {F.}~\bibnamefont {Lantelme}}, \
  and\ \bibinfo {author} {\bibfnamefont {H.}~\bibnamefont {Friedman}},\
  }\href@noop {} {\bibfield  {journal} {\bibinfo  {journal} {Journal of
  Chemical Physics}\ }\textbf {\bibinfo {volume} {66}},\ \bibinfo {pages}
  {3039} (\bibinfo {year} {1977})}\BibitemShut {NoStop}%
\bibitem [{\citenamefont {Helfand}(1978)}]{bib8}%
  \BibitemOpen
  \bibfield  {author} {\bibinfo {author} {\bibfnamefont {E.}~\bibnamefont
  {Helfand}},\ }\href@noop {} {\bibfield  {journal} {\bibinfo  {journal}
  {Journal of Chemical Physics}\ }\textbf {\bibinfo {volume} {69}},\ \bibinfo
  {pages} {1010} (\bibinfo {year} {1978})}\BibitemShut {NoStop}%
\bibitem [{\citenamefont {Lomdahl}\ and\ \citenamefont {Kerr}(1985)}]{bib81}%
  \BibitemOpen
  \bibfield  {author} {\bibinfo {author} {\bibfnamefont {P.}~\bibnamefont
  {Lomdahl}}\ and\ \bibinfo {author} {\bibfnamefont {W.}~\bibnamefont {Kerr}},\
  }\href@noop {} {\bibfield  {journal} {\bibinfo  {journal} {Physical Review
  Letters}\ }\textbf {\bibinfo {volume} {55}},\ \bibinfo {pages} {1235}
  (\bibinfo {year} {1985})}\BibitemShut {NoStop}%
\bibitem [{\citenamefont {Grozema}\ \emph {et~al.}(2002)\citenamefont
  {Grozema}, \citenamefont {Siebbeles}, \citenamefont {Berlin},\ and\
  \citenamefont {Ratner}}]{bib9}%
  \BibitemOpen
  \bibfield  {author} {\bibinfo {author} {\bibfnamefont {F.}~\bibnamefont
  {Grozema}}, \bibinfo {author} {\bibfnamefont {L.}~\bibnamefont {Siebbeles}},
  \bibinfo {author} {\bibfnamefont {Y.}~\bibnamefont {Berlin}}, \ and\ \bibinfo
  {author} {\bibfnamefont {M.}~\bibnamefont {Ratner}},\ }\href@noop {}
  {\bibfield  {journal} {\bibinfo  {journal} {CHEMPHYSCHEM}\ }\textbf {\bibinfo
  {volume} {6}},\ \bibinfo {pages} {536} (\bibinfo {year} {2002})}\BibitemShut
  {NoStop}%
\bibitem [{\citenamefont {Lakhno}\ and\ \citenamefont {Fialko}(2003)}]{bib10}%
  \BibitemOpen
  \bibfield  {author} {\bibinfo {author} {\bibfnamefont {V.}~\bibnamefont
  {Lakhno}}\ and\ \bibinfo {author} {\bibfnamefont {N.}~\bibnamefont
  {Fialko}},\ }\href@noop {} {\bibfield  {journal} {\bibinfo  {journal} {JETP
  Letters}\ }\textbf {\bibinfo {volume} {78}},\ \bibinfo {pages} {336}
  (\bibinfo {year} {2003})}\BibitemShut {NoStop}%
\bibitem [{\citenamefont {Voityuk}\ \emph {et~al.}(2000)\citenamefont
  {Voityuk}, \citenamefont {Roesch}, \citenamefont {Bixon},\ and\ \citenamefont
  {Jortner}}]{bib11}%
  \BibitemOpen
  \bibfield  {author} {\bibinfo {author} {\bibfnamefont {A.}~\bibnamefont
  {Voityuk}}, \bibinfo {author} {\bibfnamefont {N.}~\bibnamefont {Roesch}},
  \bibinfo {author} {\bibfnamefont {M.}~\bibnamefont {Bixon}}, \ and\ \bibinfo
  {author} {\bibfnamefont {J.}~\bibnamefont {Jortner}},\ }\href@noop {}
  {\bibfield  {journal} {\bibinfo  {journal} {The Journal of Physical Chemistry
  B}\ }\textbf {\bibinfo {volume} {104}},\ \bibinfo {pages} {9740} (\bibinfo
  {year} {2000})}\BibitemShut {NoStop}%
\bibitem [{\citenamefont {Jortner}\ \emph {et~al.}(2002)\citenamefont
  {Jortner}, \citenamefont {Bixon}, \citenamefont {Voityuk},\ and\
  \citenamefont {Roesch}}]{bib12}%
  \BibitemOpen
  \bibfield  {author} {\bibinfo {author} {\bibfnamefont {J.}~\bibnamefont
  {Jortner}}, \bibinfo {author} {\bibfnamefont {M.}~\bibnamefont {Bixon}},
  \bibinfo {author} {\bibfnamefont {A.}~\bibnamefont {Voityuk}}, \ and\
  \bibinfo {author} {\bibfnamefont {N.}~\bibnamefont {Roesch}},\ }\href@noop {}
  {\bibfield  {journal} {\bibinfo  {journal} {The Journal of Physical Chemistry
  A}\ }\textbf {\bibinfo {volume} {106}},\ \bibinfo {pages} {7599} (\bibinfo
  {year} {2002})}\BibitemShut {NoStop}%
\bibitem [{\citenamefont {Greenside}\ and\ \citenamefont
  {Helfand}(1981)}]{bib13}%
  \BibitemOpen
  \bibfield  {author} {\bibinfo {author} {\bibfnamefont {H.}~\bibnamefont
  {Greenside}}\ and\ \bibinfo {author} {\bibfnamefont {E.}~\bibnamefont
  {Helfand}},\ }\href@noop {} {\bibfield  {journal} {\bibinfo  {journal} {Bell
  System Technical Journal}\ }\textbf {\bibinfo {volume} {60}},\ \bibinfo
  {pages} {1927} (\bibinfo {year} {1981})}\BibitemShut {NoStop}%
\bibitem [{\citenamefont {Lakhno}\ and\ \citenamefont {Fialko}(2012)}]{bib15}%
  \BibitemOpen
  \bibfield  {author} {\bibinfo {author} {\bibfnamefont {V.}~\bibnamefont
  {Lakhno}}\ and\ \bibinfo {author} {\bibfnamefont {N.}~\bibnamefont
  {Fialko}},\ }\href@noop {} {\bibfield  {journal} {\bibinfo  {journal}
  {Russian Journal of Physical Chemistry A}\ }\textbf {\bibinfo {volume}
  {86}},\ \bibinfo {pages} {832} (\bibinfo {year} {2012})}\BibitemShut
  {NoStop}%
\bibitem [{\citenamefont {Komineas}\ \emph {et~al.}(2002)\citenamefont
  {Komineas}, \citenamefont {Kalosakas},\ and\ \citenamefont {Bishop}}]{bib16}%
  \BibitemOpen
  \bibfield  {author} {\bibinfo {author} {\bibfnamefont {S.}~\bibnamefont
  {Komineas}}, \bibinfo {author} {\bibfnamefont {G.}~\bibnamefont {Kalosakas}},
  \ and\ \bibinfo {author} {\bibfnamefont {A.}~\bibnamefont {Bishop}},\
  }\href@noop {} {\bibfield  {journal} {\bibinfo  {journal} {Physical Review
  E}\ }\textbf {\bibinfo {volume} {65}},\ \bibinfo {pages} {061905} (\bibinfo
  {year} {2002})}\BibitemShut {NoStop}%
\bibitem [{\citenamefont {Peyrard}\ and\ \citenamefont
  {Bishop}(1989)}]{bib161}%
  \BibitemOpen
  \bibfield  {author} {\bibinfo {author} {\bibfnamefont {M.}~\bibnamefont
  {Peyrard}}\ and\ \bibinfo {author} {\bibfnamefont {A.}~\bibnamefont
  {Bishop}},\ }\href@noop {} {\bibfield  {journal} {\bibinfo  {journal}
  {Physical Review Letters}\ }\textbf {\bibinfo {volume} {62}},\ \bibinfo
  {pages} {2755} (\bibinfo {year} {1989})}\BibitemShut {NoStop}%
\bibitem [{\citenamefont {Basko}\ and\ \citenamefont
  {Conwell}(2002{\natexlab{a}})}]{bib17}%
  \BibitemOpen
  \bibfield  {author} {\bibinfo {author} {\bibfnamefont {D.}~\bibnamefont
  {Basko}}\ and\ \bibinfo {author} {\bibfnamefont {E.}~\bibnamefont
  {Conwell}},\ }\href@noop {} {\bibfield  {journal} {\bibinfo  {journal}
  {Physical Review Letters}\ }\textbf {\bibinfo {volume} {88}},\ \bibinfo
  {pages} {098102} (\bibinfo {year} {2002}{\natexlab{a}})}\BibitemShut
  {NoStop}%
\bibitem [{\citenamefont {Voityuk}(2005)}]{bib18}%
  \BibitemOpen
  \bibfield  {author} {\bibinfo {author} {\bibfnamefont {A.}~\bibnamefont
  {Voityuk}},\ }\href@noop {} {\bibfield  {journal} {\bibinfo  {journal}
  {Journal of Chemical Physics}\ }\textbf {\bibinfo {volume} {122}},\ \bibinfo
  {pages} {204904} (\bibinfo {year} {2005})}\BibitemShut {NoStop}%
\bibitem [{\citenamefont {Neill}\ \emph {et~al.}(2001)\citenamefont {Neill},
  \citenamefont {Parker}, \citenamefont {Plumb},\ and\ \citenamefont
  {Siebbeles}}]{ref3-1}%
  \BibitemOpen
  \bibfield  {author} {\bibinfo {author} {\bibfnamefont {P.}~\bibnamefont
  {Neill}}, \bibinfo {author} {\bibfnamefont {A.}~\bibnamefont {Parker}},
  \bibinfo {author} {\bibfnamefont {M.}~\bibnamefont {Plumb}}, \ and\ \bibinfo
  {author} {\bibfnamefont {L.}~\bibnamefont {Siebbeles}},\ }\href@noop {}
  {\bibfield  {journal} {\bibinfo  {journal} {The Journal of Physical Chemistry
  B}\ }\textbf {\bibinfo {volume} {105}},\ \bibinfo {pages} {5283} (\bibinfo
  {year} {2001})}\BibitemShut {NoStop}%
\bibitem [{\citenamefont {Su}\ \emph {et~al.}(1979)\citenamefont {Su},
  \citenamefont {Schrieffer},\ and\ \citenamefont {Heeger}}]{refer4}%
  \BibitemOpen
  \bibfield  {author} {\bibinfo {author} {\bibfnamefont {W.}~\bibnamefont
  {Su}}, \bibinfo {author} {\bibfnamefont {J.}~\bibnamefont {Schrieffer}}, \
  and\ \bibinfo {author} {\bibfnamefont {A.}~\bibnamefont {Heeger}},\
  }\href@noop {} {\bibfield  {journal} {\bibinfo  {journal} {Physical Review
  Letters}\ }\textbf {\bibinfo {volume} {42}},\ \bibinfo {pages} {1698}
  (\bibinfo {year} {1979})}\BibitemShut {NoStop}%
\bibitem [{\citenamefont {Conwell}\ and\ \citenamefont
  {Rakhmanova}(2000)}]{refer5}%
  \BibitemOpen
  \bibfield  {author} {\bibinfo {author} {\bibfnamefont {E.}~\bibnamefont
  {Conwell}}\ and\ \bibinfo {author} {\bibfnamefont {S.}~\bibnamefont
  {Rakhmanova}},\ }\href@noop {} {\bibfield  {journal} {\bibinfo  {journal}
  {Proceedings of the National Academy of Sciences USA}\ }\textbf {\bibinfo
  {volume} {97}},\ \bibinfo {pages} {4556} (\bibinfo {year}
  {2000})}\BibitemShut {NoStop}%
\bibitem [{\citenamefont {Basko}\ and\ \citenamefont
  {Conwell}(2002{\natexlab{b}})}]{refer6}%
  \BibitemOpen
  \bibfield  {author} {\bibinfo {author} {\bibfnamefont {D.}~\bibnamefont
  {Basko}}\ and\ \bibinfo {author} {\bibfnamefont {E.}~\bibnamefont
  {Conwell}},\ }\href@noop {} {\bibfield  {journal} {\bibinfo  {journal}
  {Physical Review E}\ }\textbf {\bibinfo {volume} {65}},\ \bibinfo {pages}
  {061902} (\bibinfo {year} {2002}{\natexlab{b}})}\BibitemShut {NoStop}%
\bibitem [{\citenamefont {Conwell}\ \emph {et~al.}(2005)\citenamefont
  {Conwell}, \citenamefont {Park},\ and\ \citenamefont {Choi}}]{refer7}%
  \BibitemOpen
  \bibfield  {author} {\bibinfo {author} {\bibfnamefont {E.}~\bibnamefont
  {Conwell}}, \bibinfo {author} {\bibfnamefont {J.-H.}\ \bibnamefont {Park}}, \
  and\ \bibinfo {author} {\bibfnamefont {H.-Y.}\ \bibnamefont {Choi}},\
  }\href@noop {} {\bibfield  {journal} {\bibinfo  {journal} {The Journal of
  Physical Chemistry B}\ }\textbf {\bibinfo {volume} {109}},\ \bibinfo {pages}
  {9760} (\bibinfo {year} {2005})}\BibitemShut {NoStop}%
\bibitem [{\citenamefont {Zekovi\'c}\ \emph {et~al.}(2011)\citenamefont
  {Zekovi\'c}, \citenamefont {Zdravkovi\'c},\ and\ \citenamefont
  {Ivi\'c}}]{refer71}%
  \BibitemOpen
  \bibfield  {author} {\bibinfo {author} {\bibfnamefont {S.}~\bibnamefont
  {Zekovi\'c}}, \bibinfo {author} {\bibfnamefont {S.}~\bibnamefont
  {Zdravkovi\'c}}, \ and\ \bibinfo {author} {\bibfnamefont {Z.}~\bibnamefont
  {Ivi\'c}},\ }\href@noop {} {\bibfield  {journal} {\bibinfo  {journal}
  {Journal of Physics: Conference Series}\ }\textbf {\bibinfo {volume} {329}}
  (\bibinfo {year} {2011})}\BibitemShut {NoStop}%
\bibitem [{\citenamefont {T.Koslowski}\ \emph {et~al.}()\citenamefont
  {T.Koslowski}, \citenamefont {T.Cramer},\ and\ \citenamefont
  {N.Utz}}]{refer8}%
  \BibitemOpen
  \bibfield  {author} {\bibinfo {author} {\bibnamefont {T.Koslowski}}, \bibinfo
  {author} {\bibnamefont {T.Cramer}}, \ and\ \bibinfo {author} {\bibnamefont
  {N.Utz}},\ }\enquote {\bibinfo {title} {Atomic models of biological charge
  transfer},}\ in\ \href@noop {} {\emph {\bibinfo {booktitle} {Modern Methods
  for Theoretical Physical Chemistry of Biopolymers}}},\ \bibinfo {editor}
  {edited by\ \bibinfo {editor} {\bibfnamefont {E.}~\bibnamefont {Starikov}},
  \bibinfo {editor} {\bibfnamefont {J.}~\bibnamefont {Lewis}}, \ and\ \bibinfo
  {editor} {\bibfnamefont {S.}~\bibnamefont {Tanaka}}},\ Chap.~\bibinfo
  {chapter} {23}\BibitemShut {NoStop}%
\bibitem [{\citenamefont {Koslowski}\ and\ \citenamefont {Cramer}()}]{refer9}%
  \BibitemOpen
  \bibfield  {author} {\bibinfo {author} {\bibfnamefont {T.}~\bibnamefont
  {Koslowski}}\ and\ \bibinfo {author} {\bibfnamefont {T.}~\bibnamefont
  {Cramer}},\ }\enquote {\bibinfo {title} {Atomistic models of dna charge
  transfer.}}\ in\ \href@noop {} {\emph {\bibinfo {booktitle} {Charge Migration
  in DNA: Perspectives from Physics, Chemistry, and Biology.}}},\ \bibinfo
  {editor} {edited by\ \bibinfo {editor} {\bibfnamefont {T.}~\bibnamefont
  {Chakraborty}}},\ Chap.~\bibinfo {chapter} {4}\BibitemShut {NoStop}%
\bibitem [{\citenamefont {Lakhno}\ and\ \citenamefont
  {Fialko}(2005)}]{refer10}%
  \BibitemOpen
  \bibfield  {author} {\bibinfo {author} {\bibfnamefont {V.}~\bibnamefont
  {Lakhno}}\ and\ \bibinfo {author} {\bibfnamefont {N.}~\bibnamefont
  {Fialko}},\ }\href@noop {} {\bibfield  {journal} {\bibinfo  {journal} {The
  European Physical Journal B}\ }\textbf {\bibinfo {volume} {43}} (\bibinfo
  {year} {2005})}\BibitemShut {NoStop}%
\end{thebibliography}%

\end{document}